\documentclass[prd,twocolumn,showpacs,preprintnumbers,amsmath,amssymb,citeautoscript,nofootinbib]{revtex4}
\usepackage[dvips]{epsfig}

\newcommand{\nc}[1]{\newcommand{#1}}

\nc{\its}[1]{\itshape #1 \upshape}
\nc{\mc}[3]{\multicolumn{#1}{#2}{#3}}

\nc{\bc}{\begin{center}}
\nc{\ec}{\end{center}}
\nc{\ig}[1]{\bc \includegraphics{#1} \ec}

\nc{\bo}[1]{\mbox{\boldmath \( #1 \! \! \)  \unboldmath}}
\newcommand{\beqn} {\begin{equation}}
\newcommand{\eqn} {\end{equation}}
\nc{\be}{\begin{eqnarray}}
\nc{\ee}{\end{eqnarray}}
\nc{\bew}{\begin{eqnarray*}}
\nc{\eew}{\end{eqnarray*}}
\nc{\bs}{\begin{subeqnarray}}   
\nc{\es}{\end{subeqnarray}}     
\nc{\nnn}{\nonumber \\}
\nc{\f}[2]{\frac{#1}{#2}}
\nc{\td}[2]{\f{d #1}{d #2}}
\nc{\pd}[2]{\f{\partial #1}{\partial #2}}
\nc{\suli}{\sum\limits}
\nc{\proli}{\prod\limits}
\nc{\ili}{\int\limits}
\nc{\sr}[2]{\stackrel{#1}{#2}}
\nc{\dps}{\displaystyle}
\nc{\ket}[1]{\left| #1 \right>}
\nc{\bra}[1]{\left< #1 \right|}
\nc{\bracket}[2]{\left< #1 \right| \left. \! #2 \right>}
\nc{\norm}[1]{\left\| #1 \right\|}
\nc{\lndm}[1]{\pd{^{#1} \ln{\det{M}}}{\mu^{#1}}}
\nc{\pdmm}[1]{M^{-1} \pd{^{#1} M}{\mu^{#1}}}
\nc{\pdm}{M^{-1}\pd{M}{\mu}}
\nc{\trac}[1]{\mbox{Tr}\left(#1\right)}
\nc{\hm}{\hat{m}}
\nc{\hmu}{\hat{\mu}}

\def\lsim{\raise0.3ex\hbox{$<$\kern-0.75em\raise-1.1ex\hbox{$\sim$}}}
\def\gsim{\raise0.3ex\hbox{$>$\kern-0.75em\raise-1.1ex\hbox{$\sim$}}}

\begin{document}

\title{Baryon Number, Strangeness and Electric Charge Fluctuations \\
in QCD at High Temperature}

\author{M. Cheng\footnote{current address: Lawrence Livermore National Laboratory
Livermore, CA 94550, USA.}$^{\rm a}$, P. Hegde$^{\rm b,c}$ 
C. Jung$^{\rm c}$, F. Karsch$^{\rm c,d}$, O. Kaczmarek$^{\rm d}$, E. Laermann$^{\rm d}$,\\
R. D. Mawhinney$^{\rm a}$, C. Miao$^{\rm c}$,
P. Petreczky$^{\rm c,e}$, C. Schmidt$^{\rm d}$, W. Soeldner\footnote{current address:
Gesellschaft f\"ur Schwerionenforschung, Planckstr. 1, D-64291 Darmstadt, Germany.}$^{\rm c}$ 
}

\affiliation{
$^{\rm a}$ Physics Department,Columbia University, New York, NY 10027, USA\\
$^{\rm b}$Department of Physics and Astronomy, Stony Brook
University, Stony Brook, NY 11790, USA\\
$^{\rm c}$Physics Department, Brookhaven National Laboratory,
Upton, NY 11973, USA \\
$^{\rm d}$Fakult\"at f\"ur Physik, Universit\"at Bielefeld, D-33615 Bielefeld,
Germany\\
$^{\rm e}$ RIKEN-BNL Research Center, Brookhaven National Laboratory,
Upton, NY 11973, USA \\
}

\date{\today}
\preprint{BI-TP 2008/35}

\begin{abstract}
We analyze baryon number, strangeness and electric charge 
fluctuations as well as their correlations
in QCD at high temperature. We present results obtained from 
lattice calculations performed with an improved staggered fermion
action (p4-action) at two values of the lattice cut-off
with almost physical up and down quark masses and a physical value 
for the strange quark mass. We compare these results, with an 
ideal quark gas at high temperature and a hadron resonance gas model
at low temperature.  We find that fluctuations and correlations are 
well described by the former already for temperatures about 1.5 times 
the transition temperature. At low temperature qualitative features of
the lattice results are 
quite
well described by a hadron resonance gas model.
Higher order cumulants, which become increasingly sensitive to the 
light pions, however show deviations from a resonance gas in the 
vicinity of the transition temperature.
\end{abstract}


\maketitle

\section{Introduction}
\label{intro}
Fluctuations of conserved charges, like baryon number, electric
charge and strangeness, are generally considered to be sensitive
indicators for the structure of (subsets of) a thermal medium produced
in heavy ion collisions \cite{Koch}. In fact, if at non-vanishing baryon number
a critical point exists in the QCD phase diagram, this will be
signaled by divergent fluctuations of e.g. the baryon number density \cite{CP}.

Under conditions met in current experiments at RHIC as well as in the
upcoming heavy ion experiments at the LHC the net baryon number is 
small and QCD at vanishing chemical potential provides a good 
approximation. In this region the transition from the low temperature hadronic
to the high temperature plasma regime is continuous and fluctuations 
are not expected to lead to any singular behavior. 
Nonetheless, they provide direct insight into the structure of the
thermal medium, the relevant degrees of freedom and their correlations.
Furthermore, enhanced fluctuations provide hints for nearby singularities
in the QCD phase diagram related to the chiral limit at vanishing 
net baryon number as well as for a possible critical point at physical
values of the quark masses at non-vanishing net baryon number
density \cite{Stephanov}. 

Away from criticality, {\it i.e.} under conditions met at RHIC
and LHC, indications for the existence of such critical points can 
only show up in higher
order derivatives of the QCD partition function with respect to temperature
or chemical potentials. 
Through the analysis of fluctuations of conserved charges as well as
their higher moments and correlations we thus gain insight into the
relevant degrees of freedom of the system under consideration and at the 
same time gather information on possible nearby singularities in the 
QCD phase diagram. 
 
From lattice calculations at vanishing chemical potential
it is well known that baryon number and strangeness susceptibilities
are  sensitive indicators for the transition from the low temperature
hadronic regime to the high temperature quark gluon plasma. Also in the
case of a continuous crossover transition, as is the case for QCD
with physical quark masses, the susceptibilities rise rapidly
in the transition region \cite{Gottlieb,milc_sus,Gavai,c6,levkova}. This shows
that on the scale of the temperature the (quasi)-particles carrying the 
quantum numbers under consideration ($B,\; S$) are heavy at low  
and light at high temperature.

In calculations with two light, dynamical quark degrees of freedom (2-flavor 
QCD)  
it also has been shown that baryon number and electric charge fluctuations 
increase and that their fourth moments start to show pronounced peaks in the 
transition region from low to high temperature \cite{c6,redlich}. In fact,
higher order cumulants of e.g. baryon number fluctuations, become increasingly
sensitive to the singular behavior in the vicinity of the chiral phase
transition at zero mass and vanishing chemical potential. Starting from the
$6$th order cumulant baryon number fluctuations will diverge in the
chiral limit and are expected to reflect critical behavior 
in accordance with the $O(4)$ symmetric universality of the chiral 
transition.

It also has been noted that the analysis of correlations between
different quantum numbers or flavor channels will provide insight into
the quasi-particle structure and the relevant degrees of freedom
in QCD at high temperature \cite{vkoch}. In lattice calculations this 
has been analyzed in
quenched \cite{gavaiquenched} and 2-flavor QCD \cite{Gavai2,c6,redlich}. 
Recent lattice calculations for the QCD equation of state performed with 
dynamical light and strange quark degrees of freedom \cite{MILCeos,ourEoS}
now also allow to perform a systematic analysis of these correlations among
different quantum number channels including effects arising from the 
treatment of strange quarks as dynamical degrees of freedom
\cite{levkova,miao}.

We present here results from lattice calculations of baryon number,
electric charge and strangeness fluctuations
in QCD with dynamical light and strange quark degrees of freedom\footnote{Preliminary
results of this calculation had been presented at Lattice 2007 and 2008 \cite{miao}.}. 
The results are based on calculations with an improved staggered fermion action 
(p4-action) that strongly reduces lattice cut-off effects in bulk thermodynamics 
at high temperature. The 
values of the quark masses used in this calculation are almost physical; 
the strange quark mass, $m_s$, is fixed to its physical value while the light up 
and down quark masses are taken to be degenerate and equal to $m_s/10$. This is 
about twice as large as the 
average up and down quark masses realized in nature. We obtained
results from calculations performed with two different values
of the lattice cut-off, corresponding to lattices with temporal extent
$N_\tau=4$ and $6$. This allows us to judge the magnitude of systematic effects
arising from discretization errors in our improved action calculations. 
The spatial volume has been chosen to be $V^{1/3}T=4$, which insures that finite 
volume effects are small.  

\section{Fluctuations and Correlations; computational setup}

At vanishing baryon number ($B$), electric charge ($Q$) and strangeness ($S$)
fluctuations of these quantities can be obtained by starting from the 
QCD partition function with non-zero light and strange quark chemical 
potentials, $\hmu_{u,d,s}\equiv \mu_{u,d,s}/T$.  The quark chemical 
potentials can be expressed in terms of chemical potentials for 
baryon number ($\mu_B$), strangeness ($\mu_S$) and electric charge ($\mu_Q$),
\begin{eqnarray}
\mu_u&=&\frac{1}{3}\mu_B + \frac{2}{3}\mu_Q \; , \nonumber \\
\mu_d&=&\frac{1}{3}\mu_B - \frac{1}{3}\mu_Q \; ,\nonumber \\
\mu_s&=&\frac{1}{3}\mu_B - \frac{1}{3}\mu_Q - \mu_S \; .
\label{potential}
\end{eqnarray}

Moments of charge fluctuations, $\delta N_X \equiv N_X-\langle N_X\rangle$, 
with $X=B$, $Q$ or $S$ and their correlations are then 
obtained from derivatives of the logarithm of the QCD partition function,
{\it i.e.} the pressure,  
\begin{equation}
\frac{p}{T^4} \equiv \frac{1}{VT^3}\ln Z(V,T,\mu_B,\mu_S,\mu_Q) \; ,
\label{pressure}
\end{equation}
evaluated at $\mu_{B,Q,S}=0$,
\begin{equation}
\chi_{ijk}^{BQS} = \left. \frac{\partial^{i+j+k}p/T^4}{\partial\hmu_B^i\partial\hmu_Q^j
\partial\hmu_S^k}\right|_{\mu =0} \; ,
\label{obs}
\end{equation}
with $\hat\mu_X \equiv \mu_X/T$. 
While the first derivatives, {\it i.e.} baryon number, electric charge and
strangeness densities, vanish for $\hmu_{B,Q,S}=0$, their moments 
and correlation functions with $i+j+k$ even are non-zero. The basic quantities 
we will analyze here
are the quadratic and quartic charge fluctuations\footnote{As all expectation
values have been evaluated at vanishing chemical potential, we have 
$\delta N_X\equiv N_X$.},
\begin{eqnarray}
\chi_2^X &=& \frac{1}{VT^3}\langle N_X^2\rangle \nonumber \\
\chi_4^X &=& \frac{1}{VT^3}\left(\langle N_X^4\rangle - 
3 \langle N_X^2\rangle^2\right) \;\; ,
\label{fluc}
\end{eqnarray}
and the correlations among two conserved charges,
\begin{eqnarray}
\chi^{XY}_{11} &=&  \frac{1}{VT^3} \langle N_X N_Y\rangle \; . 
\label{cor}
\end{eqnarray}

These quantities have been evaluated in the temperature interval 
$0.8 \lsim T/T_c \lsim 2.5$ on lattices of size 
$16^3\times 4$ and $24^3\times 6$, respectively. 
On the $16^3\times 4$ lattice we also calculated $6$th order 
expansion coefficients, 
\begin{eqnarray}
\chi_6^X &=& \frac{1}{VT^3}\left( \langle N_X^6\rangle - 
15 \langle N_X^4\rangle \langle N_X^2\rangle  +
30 \langle N_X^2\rangle^3 \right)
\label{chi6}
\end{eqnarray}

The gauge field configurations, that have 
been used to evaluate the above observables, had been generated previously
in calculations of the QCD equation of state \cite{ourEoS} and the transition
temperature \cite{our_Tc}.  In these
calculations the strange quark mass has been tuned close to its physical
value and the light quark masses have been chosen to be one tenth of the
strange quark mass. This corresponds to a line of constant physics on
which the kaon mass is close to its physical value and the lightest pseudo-scalar
mass\footnote{In calculations with staggered fermions flavor symmetry is 
broken at non-vanishing values of the lattice spacing $a$. As a consequence
only one of the pseudo-scalar mesons has a light mass that is 
proportional to $\sqrt{m_l}$ and vanishes in the chiral limit at fixed
$a >0$. Full chiral symmetry with the correct Goldstone pion multiplet 
is recovered only for $a\rightarrow 0$.
For a study of the remaining flavor symmetry violations with the
p4-action in the quenched case see~\cite{FlavBreak}.
For dynamical calculations discussed here we have calculated one of
the non-Goldstone pion masses at cut-off values corresponding to
the transition region of our $N_\tau=4$ and $6$ calculations.
This gives $700$~MeV and $550$~MeV, respectively.
} is about 220~MeV, 
{\it i.e.} the light quark masses used in these calculations are about a 
factor 2 larger than their physical values.
Further details on the improved gauge and
staggered fermion actions used in these calculations are given in 
\cite{ourEoS,our_Tc}. The number of gauge field configurations 
analyzed varies from about 300 at high to 1500 at low temperatures.
Subsequent configurations are separated by 10 to 60 trajectories.
The various operators contributing to $\chi_{ijk}^{BQS}$ have been 
calculated using unbiased random estimators \cite{estimators}. While
at high temperature already 50 random sources per configuration
were sufficient to get reliable estimates for these observables,
we used up to 480 random sources below the transition temperature.
Autocorrelations of the operators contributing to the quadratic and quartic
fluctuations turned out to be about 100 trajectories close to $T_c$ and to
drop quickly away from it. Errors on these fluctuations have been 
determined by means of the jack-knife procedure.
Some details on our data sample are given in Table~\ref{tab:stat}.
\begin{table}[t]
\begin{center}
\begin{tabular}{|c|r|c|r||c|r|c|r|}
\hline
\multicolumn{4}{|c||}{$N_\tau=4$} & \multicolumn{4}{|c|}{$N_\tau=6$}  \\
\hline
T[MeV] & \#Conf. & Sep. & \#r.v. & T[MeV] & \#Conf. & Sep. & \#r.v. \\
\hline \hline
176 & 1013~ & 20 & 480~ & 174 & 985~  & 10 & 400~ \\
186 & 1550~ & 30 & 480~ & 180 & 910~  & 10 & 400~ \\
191 & 1550~ & 30 & 480~ & 186 & 1043~ & 10 & 400~ \\
195 & 1550~ & 30 & 384~ & 195 & 924~  & 10 & 400~ \\
202 & 1550~ & 30 & 384~ & 201 & 873~  & 10 & 350~ \\
205 & 475~  & 60 & 384~ & 205 & 717~  & 10 & 200~ \\
219 & 264~  & 60 & 384~ & 211 & 690~  & 10 & 150~ \\
218 & 950~  & 30 & 384~ & 224 & 560~  & 10 & 150~ \\
254 & 199~  & 60 & 192~ & 238 & 670~  & 10 & 100~ \\
305 & 302~  & 60 & 96~  & 278 & 540~  & 10 & 50~  \\
444 & 618~  & 10 & 48~  & 363 & 350~  & 10 & 50~  \\
      &      &    &     & 416 & 345~  & 10 & 50~  \\
\hline
\end{tabular}
\end{center}
\caption{The data samples analyzed on lattices of temporal extent 
$N_\tau =4$ and $6$, respectively. The columns give from left to
right the temperature values, the number of configurations used in this
calculation, the number of trajectories by which these
configurations are separated and the number of random vectors used for
the evaluation of traces.}
\label{tab:stat}
\end{table}

In the following we will present our results using a physical temperature 
scale in MeV. As explained in detail in Ref.~\cite{ourEoS}, this scale has
been obtained through detailed studies of the zero temperature static quark 
potential along the line of constant physics used also for the finite
temperature calculations. This yields unambiguously the temperature in
units of the lattice scale $r_0$ that characterizes the shape of the 
potential at distance $r_0$, {\it i.e.} $r_0$ is defined through the 
relation $\left( r^2 {{\rm d}V_{\bar{q}q}(r)}/{{\rm d}r} \right)_{r=r_0} =
1.65$. Although $r_0$ is not directly accessible in experiments, its value 
is quite well known through comparison with determinations of, e.g. the 
level splitting in the bottomonium system as well as calculations of light
meson decay constants \cite{gold,gray}. For the conversion to physical units 
we have used the value $r_0 =0.469$~fm \cite{gray}.

In calculations with the improved staggered action the
transition temperature has been determined on lattices with temporal 
extent $N_\tau=4$ and $6$ for the values of quark masses used 
here \cite{our_Tc,ourEoS}. This gave 
$T_c = 204(2)$~MeV for $N_\tau=4$ and $196(3)$~~MeV for $N_\tau=6$,
respectively. 
The temperature values given in Tables I-III have been obtained from a
fit to $r_0/a$. The fitting function is given in Eq.~22 of 
Ref.~\cite{ourEoS}.
We therefore do not quote errors on individual temperature values in the
table. As discussed in Ref.~\cite{ourEoS}, 
these errors are generally below 1\%.

\begin{table}[t]
\begin{center}
\begin{tabular}{|c|c|c|c|c|} 
\hline 
 T[MeV] & $\chi_2^u$ & $\chi_2^s$ & $\chi_4^u$ & $\chi_4^s$ \\ 
\hline 
  176.0 & 0.1656(28) & 0.0922( 9) &  0.551( 38) &  0.199( 5) \\ 
  186.4 & 0.3530(41) & 0.2186(20) &  1.246(121) &  0.561(22) \\ 
  190.8 & 0.4298(50) & 0.2660(30) &  1.662( 84) &  0.722(25) \\ 
  195.4 & 0.5144(74) & 0.3190(46) &  2.199(127) &  0.968(39) \\ 
  202.4 & 0.7095(58) & 0.4464(38) &  2.315(122) &  1.189(48) \\ 
  204.8 & 0.7689(66) & 0.4918(49) &  1.762(154) &  1.117(65) \\ 
  209.6 & 0.9452(52) & 0.6577(54) &  1.140( 98) &  1.104(58) \\ 
  218.2 & 1.0040(27) & 0.7371(25) &  0.857( 29) &  1.013(25) \\ 
  253.7 & 1.0556(28) & 0.9078(25) &  0.594( 19) &  0.754(16) \\ 
  305.2 & 1.0543(21) & 0.9851(18) &  0.543( 12) &  0.631(13) \\ 
  444.4 & 1.0158( 5) & 0.9913( 5) &  0.515(  7) &  0.536( 7) \\ 
\hline 
\end{tabular} 
\begin{tabular}{|c|c|c|c|c|} 
\hline 
 T[MeV] & $\chi_2^u$ & $\chi_2^s$ & $\chi_4^u$ & $\chi_4^s$ \\ 
\hline 
 173.8 & 0.2077(77) & 0.1167(35) &  0.803(210) &  0.244(18) \\ 
 179.6 & 0.2769(85) & 0.1643(32) &  0.915(225) &  0.310(36) \\ 
 185.6 & 0.3950(81) & 0.2341(45) &  1.348(122) &  0.563(40) \\ 
 195.0 & 0.6011(116) & 0.3743(93) &  1.786(196) &  0.963(112) \\ 
 201.3 & 0.7556(80) & 0.5027(58) &  1.001( 94) &  0.843(49) \\ 
 204.6 & 0.8093(86) & 0.5630(53) &  0.935( 80) &  0.816(48) \\ 
 211.1 & 0.8714(65) & 0.6483(49) &  0.809( 87) &  0.879(66) \\ 
 224.3 & 0.9313(61) & 0.7219(62) &  0.667( 75) &  0.900(63) \\ 
 237.7 & 0.9695(33) & 0.8319(39) &  0.571( 56) &  0.731(43) \\ 
 278.4 & 0.9975(48) & 0.9439(21) &  0.576( 36) &  0.591(32) \\ 
 362.9 & 1.0181(35) & 1.0039(21) &  0.543( 22) &  0.549(11) \\ 
 415.8 & 1.0163(18) & 1.0044(30) &  0.541( 12) &  0.563(25) \\ 
\hline 
\end{tabular} 
\end{center}
\caption{The data on quadratic and quartic fluctuations of light ($u,\; d$) and
strange ($s$) quarks obtained
from calculations on $16^3\times 4$ (upper table) and $24^3\times 6$ (lower table) lattices, respectively. 
}
\label{tab:q_data4}
\end{table}

\section{Fluctuations of light and strange quark numbers}

Before entering a discussion of fluctuations of $B$, $Q$ and $S$ it
is instructive to look into fluctuations of the partonic degrees of
freedom, the light and strange quarks. It has been noticed before that
close to the transition temperature fluctuations of the heavier 
strange quarks are suppressed relative to those of the light $u$ 
or $d$ quarks \cite{Gavai2}. At higher temperatures, however,
they approach each other. In these earlier calculations, however,
the strange quark sector has not been incorporated as a dynamical
degree of freedom in the numerical calculations. Nonetheless,
the basic observation also holds true in QCD calculations with
dynamical strange quarks.

\begin{figure}[t]
\begin{center}
\begin{minipage}[c]{7.5cm}
\begin{center}
\epsfig{file=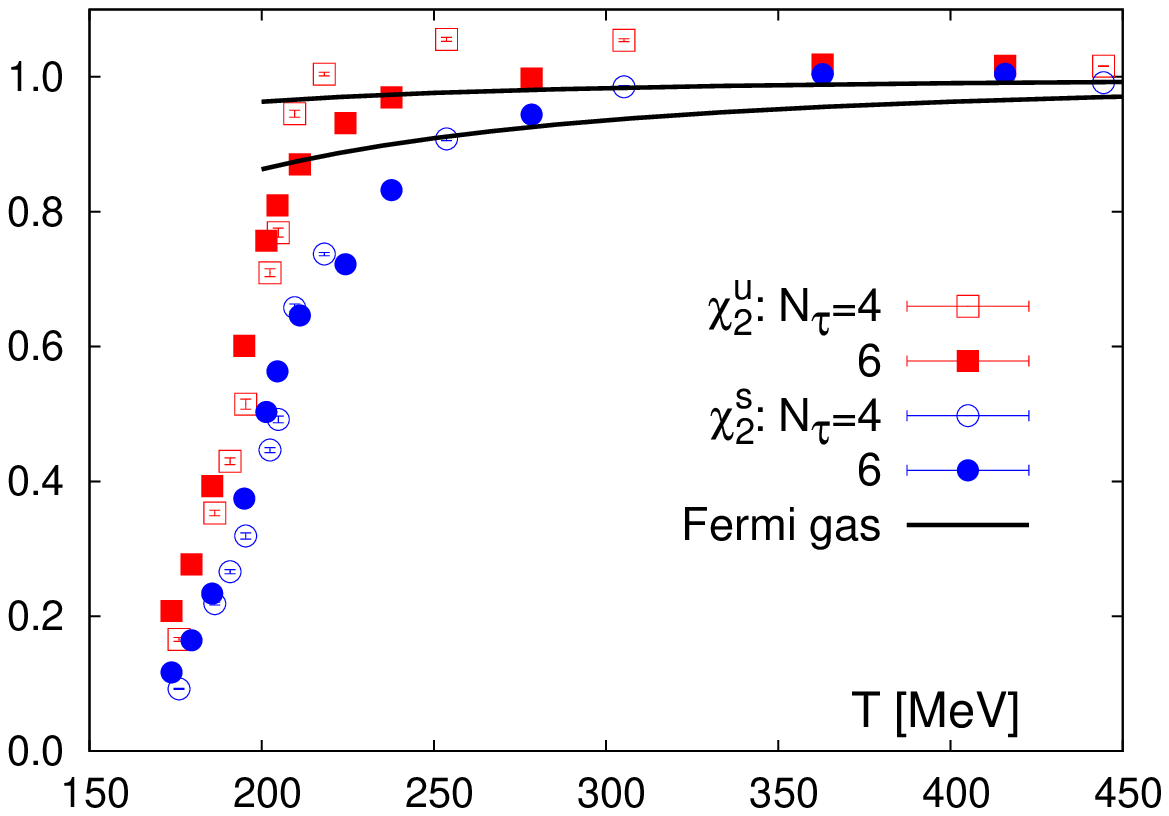, width=7.5cm}
\epsfig{file=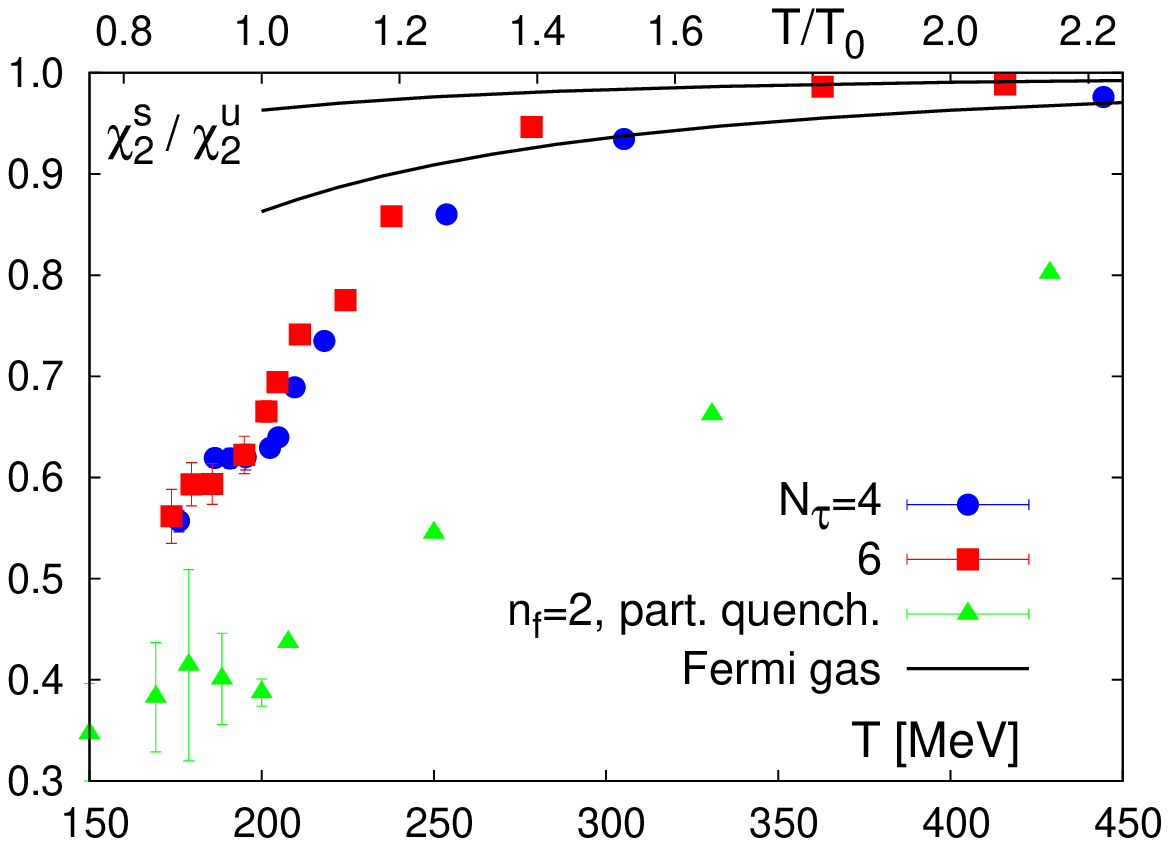, width=7.5cm}
\end{center}
\end{minipage}
\end{center}
\caption{The quadratic fluctuations of light and strange
quark number versus temperature (upper figure) and the
ratio of $s$ and $u$ quark fluctuations on lattices
of size $16^3\times 4$ and $24^3\times 6$, 
respectively. The solid lines show results for an ideal Fermi
gas with mass $m=100$~MeV and $200$~MeV,
respectively (top to bottom). 
In the lower figure we also
show partially quenched results obtained
with unimproved staggered fermions on $16^3\times 4$ lattices
\protect\cite{Gavai2}. These results have been given in
Ref.~\cite{Gavai2} as function of $T/T_c$.
This scale is shown on the upper x-axis in the lower part of the figure.
We have used the mean value of the transition temperatures on the
$N_\tau=4$ and $6$ lattices, {\it i.e.} $T_0=200$~MeV, to compare with
calculations performed with the p4-action.
}
\label{fig:c2uc2s}
\end{figure}

The difference in the behavior of light and strange quark fluctuations is 
apparent from Fig.~\ref{fig:c2uc2s}, where we show the quadratic
fluctuations of light and strange quark numbers (upper figure) and 
the ratio of these fluctuations (lower figure). The quadratic fluctuations
rapidly approach the Stefan-Boltzmann, ideal gas value, $\chi_2^{SB}=1$,
and decrease strongly in the transition region. The calculations on 
lattices with temporal extent $N_\tau=4$ and $6$ show some cut-off dependence,
which is larger for light quarks than for strange quarks. 
Nonetheless, it is apparent that the approach to the high temperature limit 
is slower for the heavier strange quarks than for the light up or down quarks. 
This is highlighted in the lower part of Fig.~\ref{fig:c2uc2s} where we
show the ratio of strange to light quark fluctuations.  
This ratio shows less cut-off dependence than $\chi_2^u$ and $\chi_2^s$ 
separately. In fact, the difference between the $N_\tau=4$ and $6$ 
results for $\chi_2^s/\chi_2^u$ is well understood in terms of the small
shift (cut-off dependence) of the transition temperature determined
for these two different lattice sizes. 

We find that $\chi_2^s/\chi_2^u$
is about $0.6$ at $T_c$ and approaches unity at about $1.7 T_c$. 
In both figures we show a band indicating the result for 
fluctuations in a non-interacting, massive Fermi gas.
As can be seen this can 
explain the smaller fluctuations of strange quarks relative to
light quarks only at temperatures larger than about $300$~MeV,
{\it i.e.} for $T\gsim 1.5\; T_c$. 
In quenched QCD  and QCD with $n_f$ massless quark flavors
the approach to the high temperature limit has also been
analyzed within HTL-resummed perturbation theory
\cite{blaizot} as well as in a straightforward
perturbative calculation performed
up to ${\cal O}(g^6\ln g)$ \cite{vuorinen}. 
In both cases
the results resemble closely properties of a 
Fermi gas with a temperature dependent mass term.
Both approaches, however, use additional 
phenomenological input to either re-sum 
next-to-leading order corrections to the 
HTL mass or fix the unknown scale for 
the ${\cal O}(g^6\ln g)$ correction.
The lattice results presented in Fig.~\ref{fig:c2uc2s}
overshoot the HTL result by about 5\%. This is of similar
magnitude as the cut-off dependence seen in the
current lattice results when comparing 
the $N_\tau=4$ and $6$ data in the temperature interval
$1.5\lsim T/T_c \lsim 2$. In view
of the large change between the leading and the next to 
leading order HTL calculations as well as the systematics 
of the cut-off dependence seen in the lattice results, 
this seems to be a reasonably good agreement. 

We note 
that the results shown in Fig.~\ref{fig:c2uc2s} are 
in agreement with calculations performed with another
improved staggered fermion action \cite{milc_sus,levkova}, the 
asqtad action. The results obtained with improved 
(p4 and asqtad) actions are, however, in contrast to 
calculations performed with the standard staggered action 
in quenched \cite{gavaiquenched} and 2-flavor \cite{Gavai2} 
QCD. These calculations led to significantly larger values
for $\chi_2$, which is understood in terms of the large 
cut-off effects in the standard discretization scheme
(see appendix). In the absence of results on larger lattices,
which would have allowed for a continuum extrapolation,
the numerical results obtained within the standard discretization
scheme \cite{gavaiquenched,Gavai2} have been normalized to the
corresponding ideal gas value evaluated also on lattices
with finite temporal extent.
It is well known that at temperatures 
a few times the transition temperature
this procedure, which is correct in the limit of
infinite temperature, over-estimates the 
actual cut-off distortion of numerical results.
The normalized values thus end up to be substantially smaller
than the results obtained with improved staggered fermion
actions for which an ad-hoc normalization has not been
performed. We discuss cut-off effects and their quark mass
dependence for the standard and p4-action in
somewhat more detail in the appendix.

The situation is different for 
the ratio $\chi_2^s/\chi_2^u$ shown in the lower part of 
Fig.~\ref{fig:c2uc2s}. Here unknown normalization
factors drop out. Nonetheless, also in this case results
obtained in calculations with improved actions 
are significantly larger than those obtained with
a standard action. This difference may be due to the
fact that in Ref.~\cite{Gavai2} the strange quark 
sector has only been treated in the quenched approximation.

\section{Fluctuation of conserved charges}

\begin{table}[t]
\begin{center}
\begin{tabular}{|c|c|c|c|c|} 
\hline 
 T[MeV] & $\chi_2^B$ & $\chi_2^Q$ & $\chi_4^B$ & $\chi_4^Q$ \\ 
\hline 
  176.0 & 0.0301( 9) & 0.1229( 8) &  0.028(12) &  0.189( 5) \\ 
  186.4 & 0.0831(25) & 0.2433(17) &  0.083(16) &  0.415(17) \\ 
  190.8 & 0.1061(19) & 0.2903(29) &  0.137(18) &  0.521(26) \\ 
  195.4 & 0.1328(32) & 0.3412(43) &  0.193(15) &  0.682(30) \\ 
  202.4 & 0.1931(17) & 0.4601(35) &  0.193(36) &  0.744(35) \\ 
  204.8 & 0.2133(33) & 0.4961(39) &  0.111(25) &  0.599(41) \\ 
  209.6 & 0.2694(23) & 0.6154(38) &  0.072(14) &  0.365(43) \\ 
  218.2 & 0.2978( 5) & 0.6480(18) &  0.060( 5) &  0.259(11) \\ 
  253.7 & 0.3313(18) & 0.6918(12) &  0.034( 6) &  0.159( 6) \\ 
  305.2 & 0.3398(10) & 0.6992(11) &  0.023( 5) &  0.143( 3) \\ 
  444.4 & 0.3343( 3) & 0.6760( 5) &  0.019( 2) &  0.127( 1) \\ 
\hline 
\end{tabular} 
\begin{tabular}{|c|c|c|c|c|} 
\hline 
 T[MeV] & $\chi_2^B$ & $\chi_2^Q$ & $\chi_4^B$ & $\chi_4^Q$ \\ 
\hline 
 173.8 & 0.0395(36) & 0.1517(30) &  0.094(56) &  0.215(10) \\ 
 179.6 & 0.0581(48) & 0.1980(34) &  0.101(54) &  0.260(36) \\ 
 185.6 & 0.0928(30) & 0.2700(47) &  0.102(32) &  0.435(34) \\ 
 195.0 & 0.1600(37) & 0.3930(78) &  0.142(30) &  0.577(69) \\ 
 201.3 & 0.2088(31) & 0.4928(46) &  0.047(18) &  0.346(29) \\ 
 204.6 & 0.2299(41) & 0.5266(38) &  0.091(15) &  0.270(20) \\ 
 211.1 & 0.2528(28) & 0.5723(32) &  0.067(22) &  0.237(18) \\ 
 224.3 & 0.2794(21) & 0.6058(37) &  0.036(17) &  0.218(19) \\ 
 237.7 & 0.3000(20) & 0.6388(15) &  0.043(23) &  0.156( 7) \\ 
 278.4 & 0.3232(22) & 0.6637(16) &  0.029( 6) &  0.139( 7) \\ 
 362.9 & 0.3354(11) & 0.6798(17) &  0.021( 7) &  0.135( 5) \\ 
 415.8 & 0.3362(10) & 0.6777(12) &  0.023( 4) &  0.131( 2) \\ 
\hline 
\end{tabular} 
\end{center}
\caption{The data on quadratic and quartic fluctuations obtained
from calculations on $16^3\times 4$ (upper table) and  $24^3\times 6$ 
(lower table) lattices, respectively. 
}
\label{tab:data4}
\end{table}

\begin{figure}[t]
\begin{center}
\begin{minipage}[c]{7.5cm}
\begin{center}
\epsfig{file=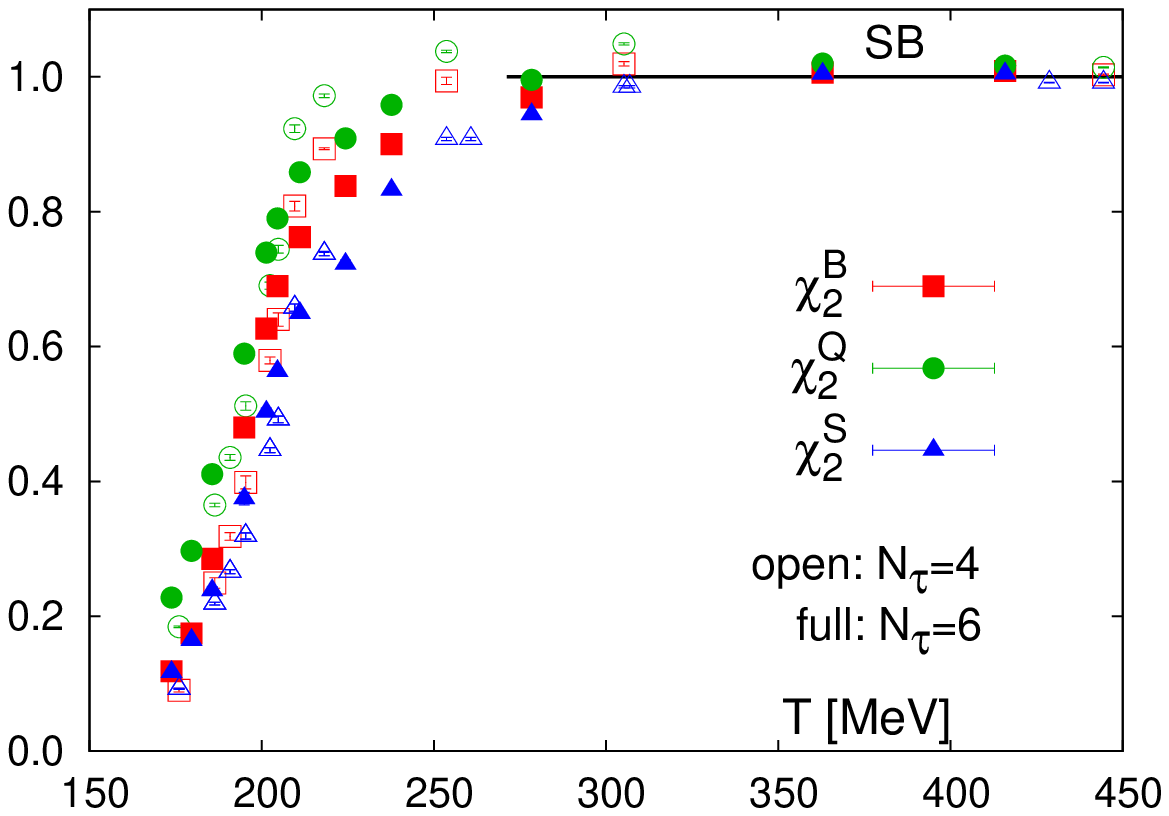, width=7.5cm}
\epsfig{file=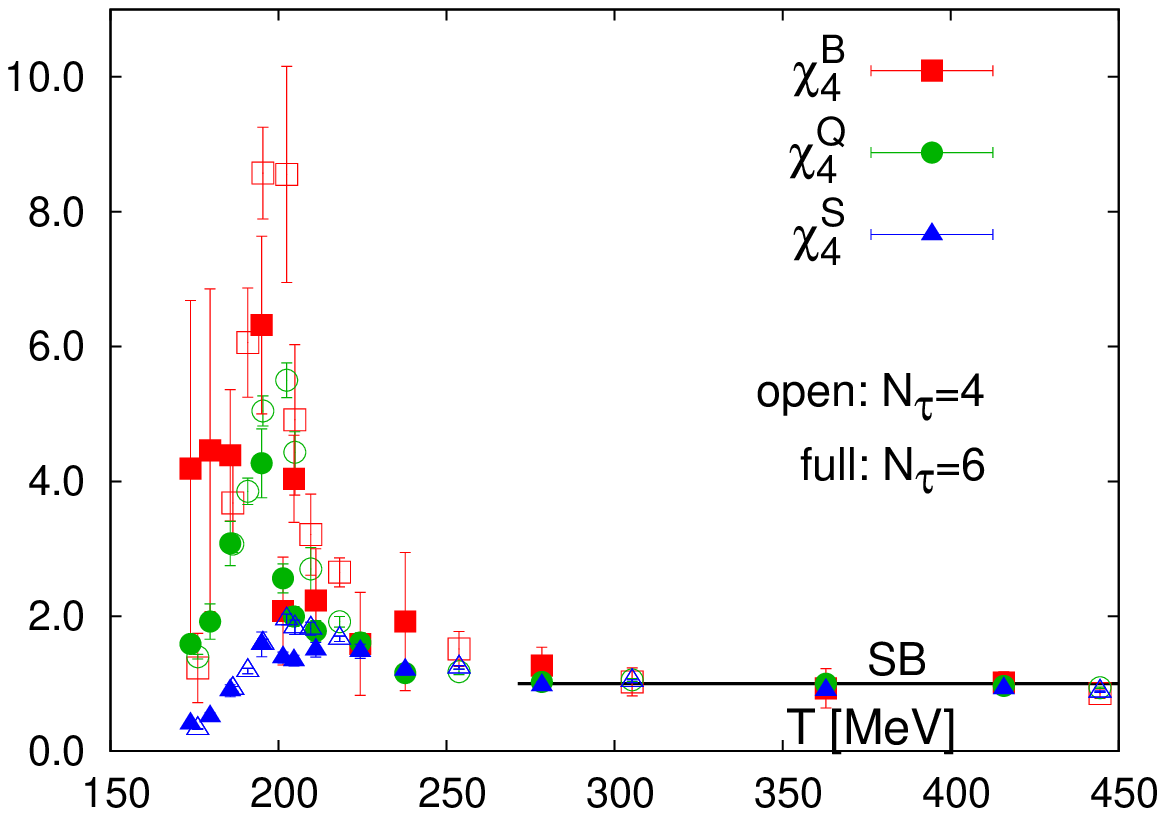, width=7.5cm}
\end{center}
\end{minipage}
\end{center}
\caption{Quadratic and quartic fluctuations of baryon number,
electric charge and strangeness. All quantities have been 
normalized to the corresponding free quark gas values, $\chi_n^{X,SB}$
given in Table~\protect\ref{tab:SB}.
}
\label{fig:fluc}
\end{figure}

In Fig.~\ref{fig:fluc} we show results for quadratic
($\chi_2^X$) and quartic ($\chi_4^X$) fluctuations with $X=B,\; Q$ and $S$.
As can be seen, in all cases the quadratic fluctuations rise rapidly
in the transition region where the quartic fluctuations show a maximum.
This maximum is most pronounced for the baryon number fluctuations but is
still visible also in fluctuations of the strangeness number.  

We note that results obtained on lattices with temporal extent
$N_\tau =6$ are in good agreement with those obtained on the 
coarser $N_\tau=4$ lattice. The slight shift towards smaller 
temperatures, visible for the $N_\tau=6$ data
relative to the $N_\tau=4$ data, is consistent with findings for 
the equation of state, e.g. the trace anomaly $(\epsilon -3p)/T^4$,
and also reflects the shift in the transition temperature
observed when comparing the location of cusp in the chiral 
susceptibility \cite{our_Tc}. The generic form of this temperature
dependence, a smooth crossover for quadratic fluctuations and 
a peak in quartic fluctuations, is in fact expected to occur
in the vicinity of the chiral phase transition of QCD. At 
vanishing chemical potential and for vanishing quark mass
the singular behavior of fluctuation observables like, e.g.
the baryon number cumulants, is expected to be controlled
by the singular part of the free energy that has the same 
universal structure as that of a three dimensional $O(4)$
spin model\footnote{As chiral symmetry is explicitly broken
in numerical calculations with staggered fermions, the 
relevant symmetry group is, in fact, expected to be 
$O(2)$.}. This singular part of the free energy, $f_s$, is 
controlled by a reduced 'temperature' $t$ that is a function 
of temperature as well as the quark chemical potentials 
$\mu_X$ \cite{Hatta,karsch07}. The latter add in even combinations
to the reduced temperature in order to respect 
charge symmetry at $\mu_X=0$.
Baryon number cumulants are thus expected to scale like,
\begin{equation}
\chi_{2n}^B \sim \left| \frac{T-T_c}{T_c}\right|^{2-n-\alpha}
\label{scaling}
\end{equation}
with $\alpha \simeq - 0.25\; [-0.015]$ \cite{Onscaling}
denoting the critical 
exponent characterizing the non-analytic structure of
the specific heat in 3-d, $O(4)\; [O(2)]$ symmetric 
spin models. Starting at $6$th order the baryon number
cumulant thus will diverge at $T_c$ in the chiral limit.
It has been argued in \cite{c6} that this singularity 
will also show up in cumulants of the electric charge.

On lattices with temporal extent $N_\tau=4$ and $6$ and for the 
values of quark masses used here the transition temperature is
close to $200$~MeV. From  Fig.~\ref{fig:fluc} we thus conclude that
at temperatures of about $1.5T_c$ and larger quadratic and 
quartic cumulants of the fluctuations of $B$, $Q$ and $S$ are 
close to those of an ideal, massless quark gas, for 
which the pressure is given by
\begin{eqnarray}
\frac{p^{SB}}{T^4} &=&  \sum_{f=u,d,s} \left[\frac{7 \pi^2}{60} +
\frac{1}{2}  \left(\frac{\mu_f}{T}\right)^2 
+ \frac{1}{4 \pi^2} \left(\frac{\mu_f}{T}\right)^4 
\right] \; .
\label{eq:free}
\end{eqnarray}
Using the relations given in Eq.~\ref{potential} one easily derives
the corresponding ideal gas values for quadratic and quartic
fluctuations of conserved charges. These are summarized in 
Table~\ref{tab:SB}.
\begin{table}[t]
\begin{center}
\begin{tabular}{|c|c|c|c|}
\hline
X&$B$&$Q$&$S$\\
\hline
~$\chi_2^{X,SB}$~ & $1/3$ & $2/3$ & $1$ \\
~$\chi_4^{X,SB}$~ & $2/9\pi^2$ & $4/3\pi^2$ & $6/\pi^2$ \\
\hline
\end{tabular}
\end{center}
\caption{Ideal gas values for quadratic ($\chi_2$)and 
quartic ($\chi_4$)
fluctuations of baryon number ($B$), electric charge ($Q$)
and strangeness ($S$).
}
\label{tab:SB}
\end{table}

The upper part of Fig.~\ref{fig:fluc} shows that over a wide temperature 
range quadratic fluctuations of strangeness are suppressed relative to 
those of baryon number and charge, which
receive contributions also from fluctuations of the light $u$ and 
$d$ quarks. Only for temperatures $T\gsim 1.7 T_c$ do the 
fluctuations of all three charges agree. This resembles the pattern
of the relative strength of light and strange quark fluctuations
seen in Fig.~\ref{fig:c2uc2s}. 

\begin{figure}[t]
\begin{center}
\begin{minipage}[c]{7.5cm}
\begin{center}
\epsfig{file=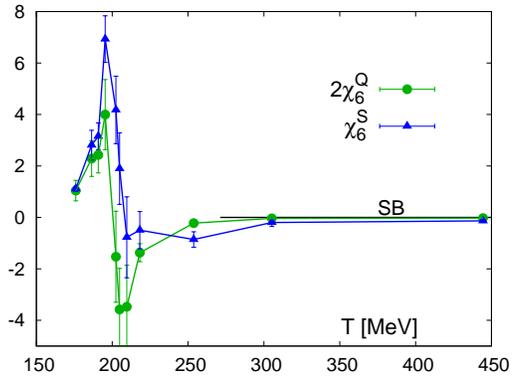, width=7.5cm}
\end{center}
\end{minipage}
\end{center}
\caption{The $6$th order cumulant of 
electric charge and strangeness fluctuations evaluated on lattices 
of size $16^3\times 4$.
}
\label{fig:chi6}
\end{figure}

On the coarser $N_\tau=4$ lattices we also have calculated the $6$th
order cumulants, $\chi_6^X$, which vanish in the infinite temperature,
ideal gas limit as well as in leading order perturbation theory \cite{blaizot}. 
For strangeness and charge fluctuations we show results in 
Fig.~\ref{fig:chi6}. Note that these $6$th
order cumulants change sign at $T\simeq T_c$. Below $T_c$ they rise 
rapidly and reach a maximum at $T\simeq 0.95 T_c$; 
they approach zero from below and almost vanish for $T\gsim 1.5 T_c$. 

\section{Ratios of cumulants}

At low temperatures hadrons are the relevant degrees of 
freedom. The hadron resonance gas (HRG) model has been shown to provide a 
good description of thermal conditions at freeze-out \cite{cleymans,pbm}. 
Also fluctuations
of the thermal medium have been successfully described in the framework of 
a HRG model \cite{tawfik}. Nonetheless, in a HRG model cumulants are monotonically 
rising functions of the temperature, while the $6$th order cumulants
change sign at $T_c$. This indicates that a straightforward HRG model
has to break down in the vicinity of the transition temperature. 

\begin{figure}[t]
\begin{center}
\begin{minipage}[c]{7.5cm}
\begin{center}
\epsfig{file=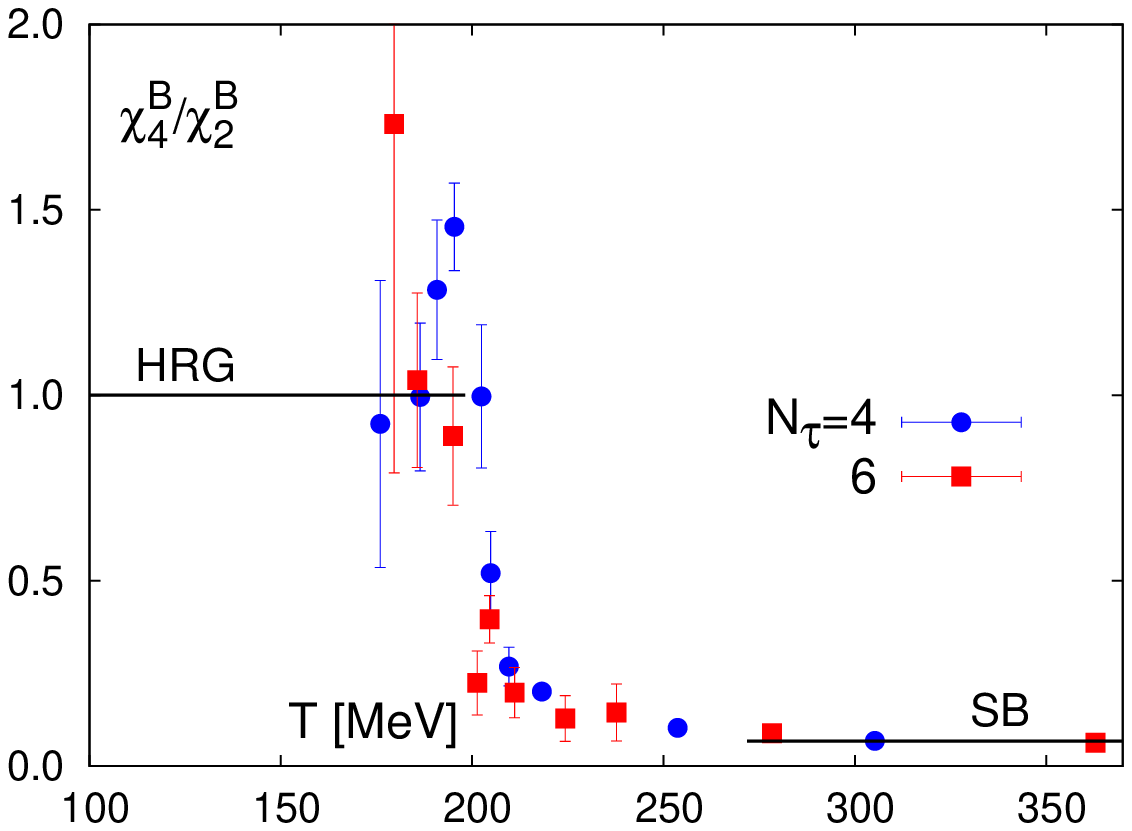, width=7.5cm}
\epsfig{file=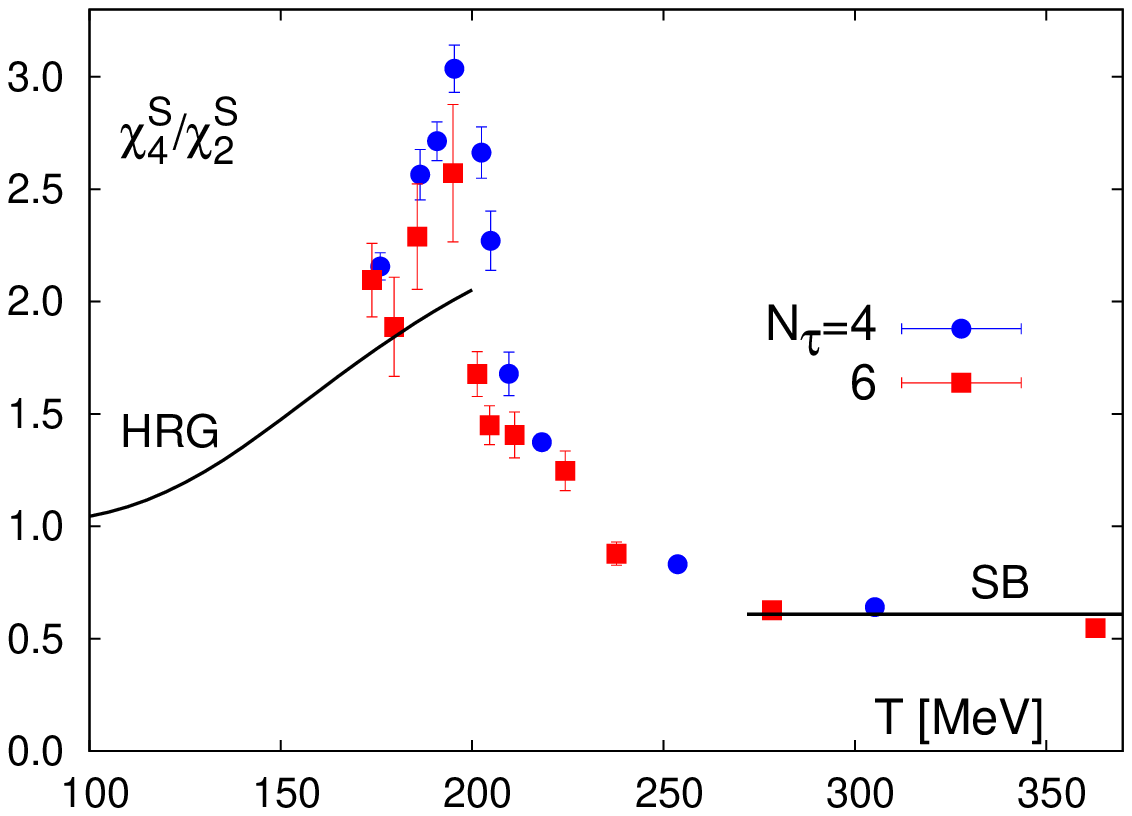, width=7.5cm}
\epsfig{file=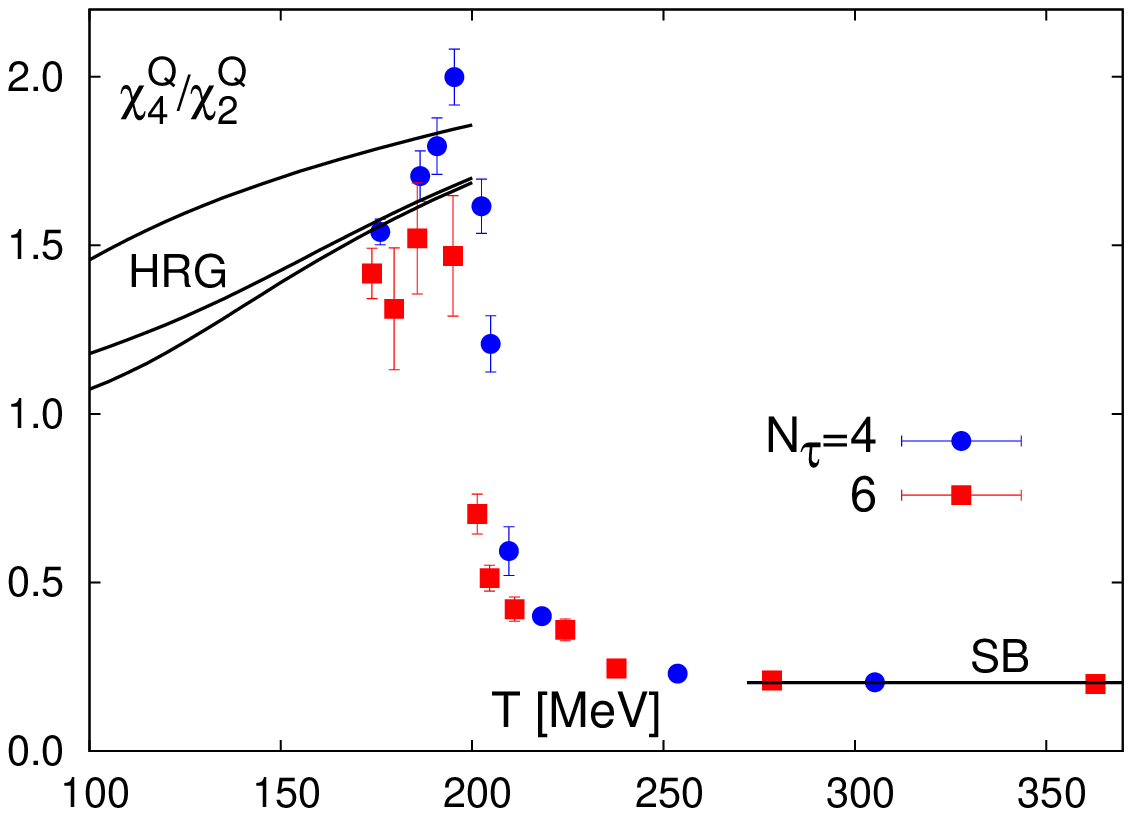, width=7.5cm}
\end{center}
\end{minipage}
\end{center}
\caption{The ratio of fourth and second order cumulants of baryon number (top), 
strangeness (middle) and electric charge (bottom) fluctuations. In the latter
case we show curves for a HRG model calculated with physical pion masses (upper
curve), pions of mass $220$~MeV (middle) and infinitely heavy pions (lower 
curve).}
\label{fig:kurtosis}
\end{figure} 

The partition
function of the hadron resonance gas can be split into mesonic and
baryonic contributions,
\begin{eqnarray}
p^{HRG}/T^4 \hspace{-2mm}
&=&\frac{1}{VT^3}\sum_{i\in\;mesons}\hspace{-3mm} 
\ln{\cal Z}^{M}_{m_i}(T,V,\mu_B,\mu_Q,\mu_S)
\nonumber \\
&&\hspace{-3mm} +\frac{1}{VT^3}
\sum_{i\in\;baryons}\hspace{-3mm} \ln{\cal Z}^{B}_{m_i}(T,V,\mu_B,\mu_Q,\mu_S)\; ,
\label{eq:ZHRG}
\end{eqnarray}
where
\begin{equation}
\ln{\cal Z}^{M/B}_{m_i}
=\mp \frac{V d_i}{{2\pi^2}} \int_0^\infty dk k^2
\ln(1\mp z_ie^{-\varepsilon_i/T}) \quad ,
\label{eq:ZMB}
\end{equation}
with energies $\varepsilon_i=\sqrt{k^2+m_i^2}$, degeneracy 
factors $d_i$ and fugacities
\begin{equation}
z_i=\exp\left((B_i\mu_B+Q_i\mu_Q+S_i\mu_S)/T\right) \; .
\label{eq:fuga}
\end{equation}
The HRG model has been compared to lattice
results previously \cite{tawfik}. 
In these earlier comparisons with lattice 
calculations, which had to be performed with rather
large quark masses and without dynamical strange 
quarks the masses contributing to the HRG model were 
adjusted and strange contributions were suppressed.
Here we no longer follow this strategy but compare
directly the lattice results with an HRG model that
is based on the experimentally observed spectrum.
We use the same HRG model
as it also is used in heavy ion phenomenology and the analysis 
of freeze-out conditions \cite{cleymans,pbm}.
This ansatz for the HRG includes all meson and baryon masses from
the particle data book with masses $m_i\le 2.5$~GeV that are 
characterized at least as '4 star states'.

Also on the lattice with temporal extent $N_\tau=6$ calculations 
at low temperature are still performed on quite coarse lattices.
We thus cannot
expect to reproduce details of the hadron spectrum well
under these conditions. 
Nevertheless, we feel that it is now most appropriate to compare
lattice results directly with the unmodified continuum version of
the HRG model that is commonly used in the phenomenological treatment
of QCD thermodynamics. By improving lattice calculations systematically
one will then be able to judge to what extent the HRG model does
describe the thermodynamics of QCD in the hadronic phase.
In fact, the current calculations of the 
QCD equation of state, e.g.  the pressure and trace anomaly,
$(\epsilon -3p)/T^4$, show deviations from the HRG at 
temperatures below the transition region. 
This is still the case in calculations on lattices with temporal
extent $N_\tau=8$, {\it i.e.} closer to the continuum limit 
\cite{karsch,Rgupta}. We thus concentrate
here on observables which are less sensitive to details of 
the hadron mass spectrum and emphasize the charges of the
relevant degrees of freedom contributing to the fluctuations.
These are in general ratios of cumulants
of $N_B, \; N_Q$, and $N_S$. In fact, in the framework of 
a HRG model it is easy to
convince oneself that the ratio of fourth and second order
cumulants of baryon number fluctuations is completely
independent of the actual value of hadron masses;
$(\chi_4^B/\chi_2^B)_{HRG} = 1$ if all hadrons are heavy
on the scale of the temperature. In that case the fugacity 
expansion of Eq.~\ref{eq:ZMB},
\begin{eqnarray}
\ln Z_m^B &=& \frac{VT^3}{\pi^2} \left( \frac{m}{T} \right)^2
\sum_{\ell=1}^{\infty} (-1)^{\ell+1}\,\ell^{-2}\, K_2(\ell m/T)\,
\nonumber \\
&&\times \cosh (\ell \mu_B/T)\quad ,
\label{bessel}
\end{eqnarray}
is well approximated by its leading order term, {\it i.e.} the
Boltzmann approximation. In this case the baryonic contribution
to the pressure of a HRG is proportional to $\cosh(\mu_B/T)$
and the ratio $\chi_4^B/\chi_2^B$ thus becomes independent of
the hadron mass spectrum.
This is 
reasonably
well reproduced by the lattice results
shown in Fig.~\ref{fig:kurtosis}(top). Note, however,
that in the chiral limit it is expected that the cusp
in $\chi_4^B$ (Fig.~\ref{fig:fluc}) is expected to become
more pronounced and thus more
prominent also in the ratio $\chi_4^B/\chi_2^B$. How 
strong this effect will be requires more detailed
studies and also a better control over the continuum 
limit. Model calculations yield quite a different strength
for the peak in $\chi_4^B/\chi_2^B$ \cite{friman}.

Even within the Boltzmann approximation the structure of
a HRG is, however, more complicated in the strange and/or 
electrically charged sectors. In these cases 
multiply strange hadrons or hadrons with charge $Q=2$
contribute to the HRG. This enhances 
quartic fluctuations relative to quadratic ones
and leads to a deviation of cumulant ratios from unity.  
This qualitative feature is
indeed seen in the results obtained for $\chi_4^S/\chi_2^S$
and $\chi_4^Q/\chi_2^Q$ shown in  Fig.~\ref{fig:kurtosis}, 
respectively. 

\begin{figure}[t]
\begin{center}
\begin{minipage}[c]{7.5cm}
\begin{center}
\epsfig{file=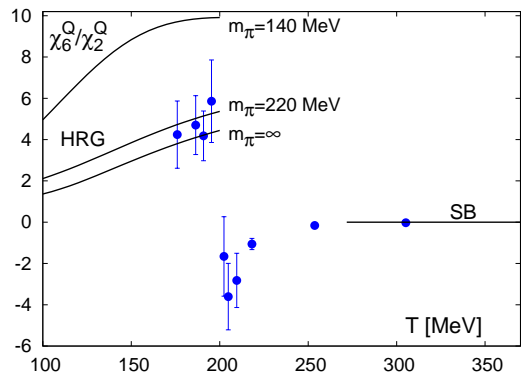, width=7.5cm}
\end{center}
\end{minipage}
\end{center}
\caption{The ratio of $6$th and $2$nd order cumulants of
electric charge fluctuations evaluated on lattices
of size $16^3\times 4$.
}
\label{fig:chi62Q}
\end{figure}

We note that the lightest hadrons, the pions, only contribute
to cumulants of electric charge fluctuations. For these light
states the Boltzmann approximation is not sufficient at 
temperatures close to $T_c$. We thus have used the complete
bosonic fugacity sum in the pion sector. These light hadronic
states are also not well taken into account in current lattice
calculations which still are being performed with light
quark masses that are about a factor two larger than the
physical up and down quark masses. Moreover, the pion spectrum
is distorted in calculations with staggered fermions
due to cut-off effects that explicitly break
flavor
symmetry at non-vanishing lattice spacing (see footnote 3).
In the case of electric charge fluctuations we therefore
have analyzed in more detail the contribution of pions to
the fluctuations in a HRG. In Fig.~\ref{fig:kurtosis}(bottom) we show
results for HRG model calculations with physical pion masses,
with a pion mass of $220$~MeV, which corresponds to the lightest
pseudo-scalar state formed with the quark mass values currently
used in our calculations, and without the pion sector, {\it i.e.} 
for infinitely heavy pions. Although the largest part of the 
electric charge fluctuations arises from heavier hadrons,
the ratio of cumulants is sensitive to the pion sector.
Already pion masses that are 50\% larger than the physical value
drastically reduce the contribution of the pion sector. This 
effect is even more significant in higher order cumulants. 
In the $6$th order cumulant half of the fluctuations can
be attributed to the light pion sector. This is obvious from
Fig.~\ref{fig:chi62Q} where we show the ratio of $6$th and $2$nd 
order cumulants for electric charge fluctuations. In a HRG model 
the ratio of cumulants calculated
without the light pion sector is a factor two smaller than 
that calculated with the physical pions included. In the low 
temperature regime the lattice calculations are consistent
with the former. We thus conclude that lighter quark masses
and calculations closer to the continuum limit are needed
to correctly represent in lattice calculations higher order 
cumulants that are sensitive to the light hadron sector.
Irrespective of this we find, however, that the 
occurrence of maxima in $\chi_6^X$ close but below $T_c$ signal
the breakdown of the HRG model close to $T_c$. The ratio
$\chi_6^Q/\chi_2^Q$ starts dropping below $T_c$ and is 
consistent with zero at $T_c$.

A similar behavior is found for higher order cumulants 
of strangeness fluctuations.  We find that for
both values of the lattice cut-off the ratio $\chi_4^S/\chi_2^S$
overshoots the HRG values in the transition region and this 
also holds true for the ratio $\chi_6^S/\chi_2^S$ evaluated on 
the coarse $16^3\times 4$ lattices.
Of course, this requires confirmation through calculations
with lighter quark masses on lattices closer to the continuum
limit. It may suggest that the contribution of even heavier,
experimentally not well-established multiple strange hadrons, 
which are not included in the
current version of the HRG model, is of importance in the 
transition region. In general we find, however, that the HRG
model gives a fairly good description of cumulants of 
the fluctuations of conserved charges up to temperatures
close to the transition temperature.

\section{Correlations among conserved charges}

\begin{table}[t]
\begin{center}
\begin{tabular}{|c|c|c|c|} 
\hline 
 T[MeV] & $\chi_{11}^{BQ}$ & $\chi_{11}^{BS}$ & $\chi_{11}^{QS}$ \\ 
\hline 
  176.0 & 0.00630(35) & -0.0175( 9) & 0.0374( 2) \\ 
  186.4 & 0.01332(40) & -0.0564(14) & 0.0811( 5) \\ 
  190.8 & 0.01670(26) & -0.0727(13) & 0.0966( 9) \\ 
  195.4 & 0.02023(43) & -0.0924(19) & 0.1133(14) \\ 
  202.4 & 0.02811(32) & -0.1369(14) & 0.1548(13) \\ 
  204.8 & 0.02959(41) & -0.1544(20) & 0.1687(15) \\ 
  209.6 & 0.03024(58) & -0.2089(19) & 0.2244(19) \\ 
  218.2 & 0.02912(17) & -0.2395( 8) & 0.2488( 9) \\ 
  253.7 & 0.01625(12) & -0.2988(16) & 0.3045( 6) \\ 
  305.2 & 0.00765( 6) & -0.3245(11) & 0.3303( 5) \\ 
  444.4 & 0.00272( 1) & -0.3289( 4) & 0.3312( 3) \\ 
\hline 
\end{tabular} 
\begin{tabular}{|c|c|c|c|c|} 
\hline 
 T[MeV] & $\chi_{11}^{BQ}$ & $\chi_{11}^{BS}$ & $\chi_{11}^{QS}$ \\ 
\hline 
 173.8 & 0.00830(88) & -0.0228(23) & 0.0469( 8) \\ 
 179.6 & 0.01040(120)& -0.0373(26) & 0.0635(14) \\ 
 185.6 & 0.01611(74) & -0.0606(18) & 0.0868(18) \\ 
 195.0 & 0.02406(49) & -0.1119(33) & 0.1312(31) \\ 
 201.3 & 0.02697(72) & -0.1549(24) & 0.1739(18) \\ 
 204.6 & 0.02641(72) & -0.1771(31) & 0.1930(13) \\ 
 211.1 & 0.02312(67) & -0.2066(25) & 0.2208(16) \\ 
 224.3 & 0.02309(90) & -0.2332(25) & 0.2444(20) \\ 
 237.7 & 0.01529(66) & -0.2695(23) & 0.2812(10) \\ 
 278.4 & 0.00529(77) & -0.3126(10) & 0.3157( 8) \\ 
 362.9 & 0.00145(63) & -0.3325( 8) & 0.3357(10) \\ 
 415.8 & 0.00122(34) & -0.3338(13) & 0.3353(12) \\ 
\hline 
\end{tabular} 
\end{center}
\caption{The data on correlations between different quantum number
fluctuations obtained from calculations on lattices with temporal
extent $N_\tau =4$ (upper table) and $6$ (lower table), respectively.
}
\label{tab:cor_data}
\end{table}

\begin{figure}[t]
\begin{center}
\begin{minipage}[c]{7.5cm}
\begin{center}
\epsfig{file=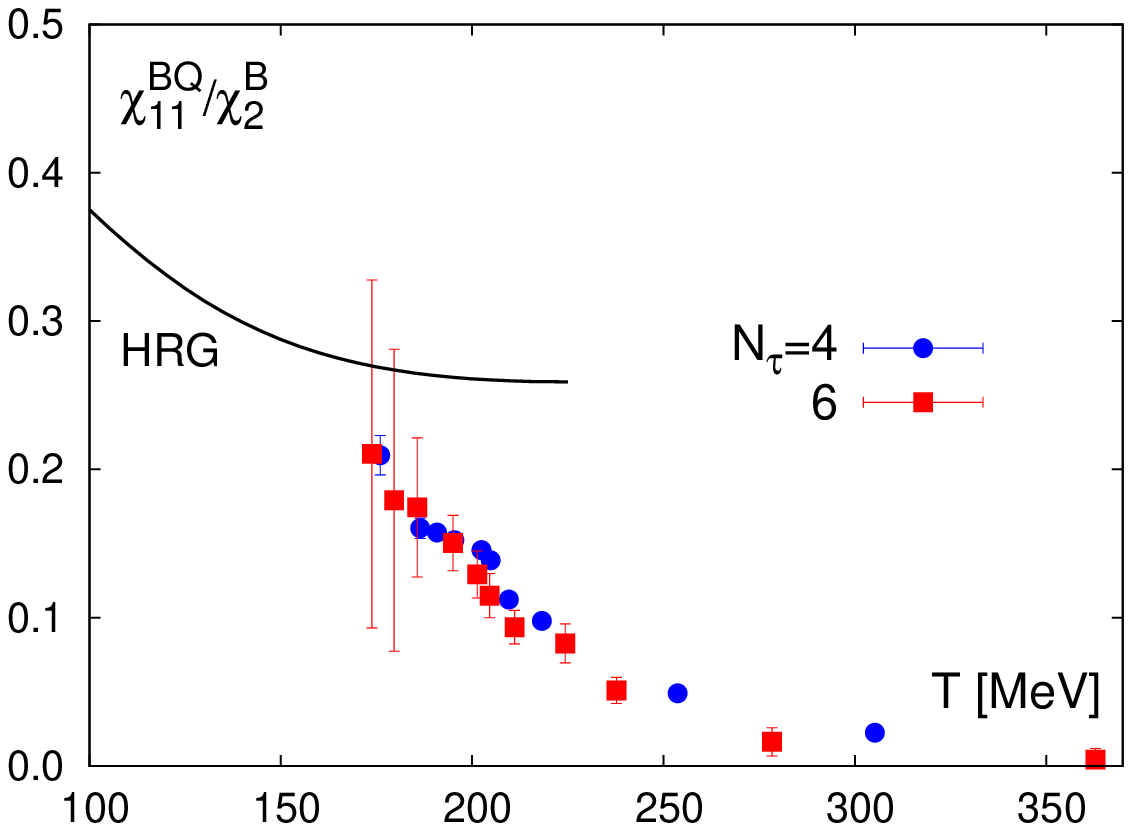, width=7.5cm}
\epsfig{file=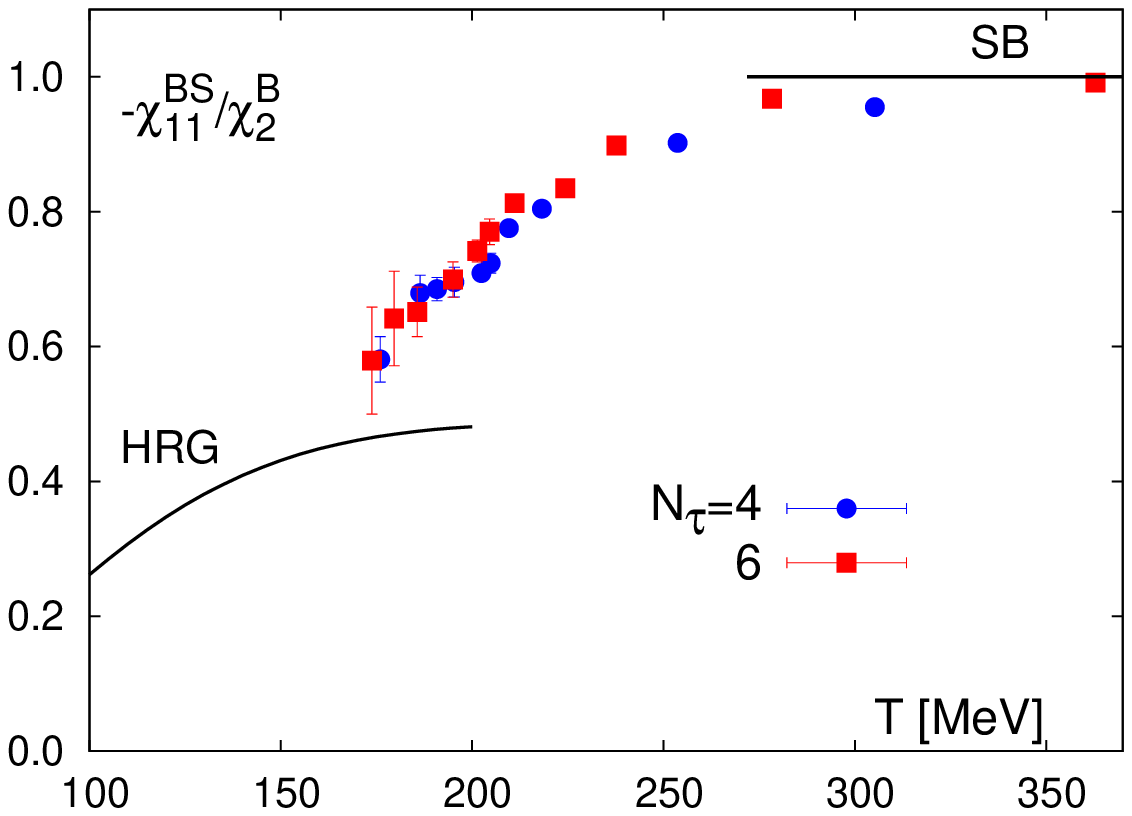, width=7.5cm}
\end{center}
\end{minipage}
\end{center}
\caption{Correlations of electric charge and strangeness with
baryon number normalized to quadratic fluctuations of baryon 
number.
}
\label{fig:cor}
\end{figure}

\begin{figure}[t]
\begin{center}
\begin{minipage}[c]{7.5cm}
\begin{center}
\epsfig{file=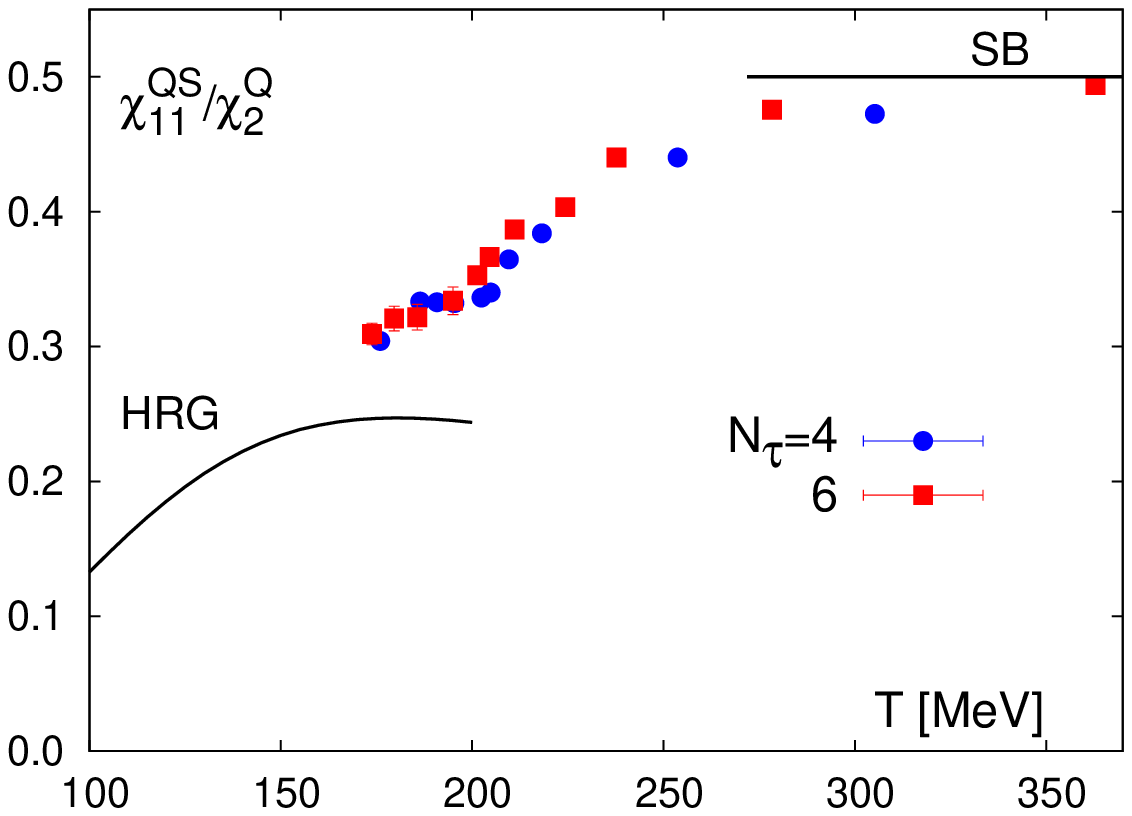, width=7.5cm}
\epsfig{file=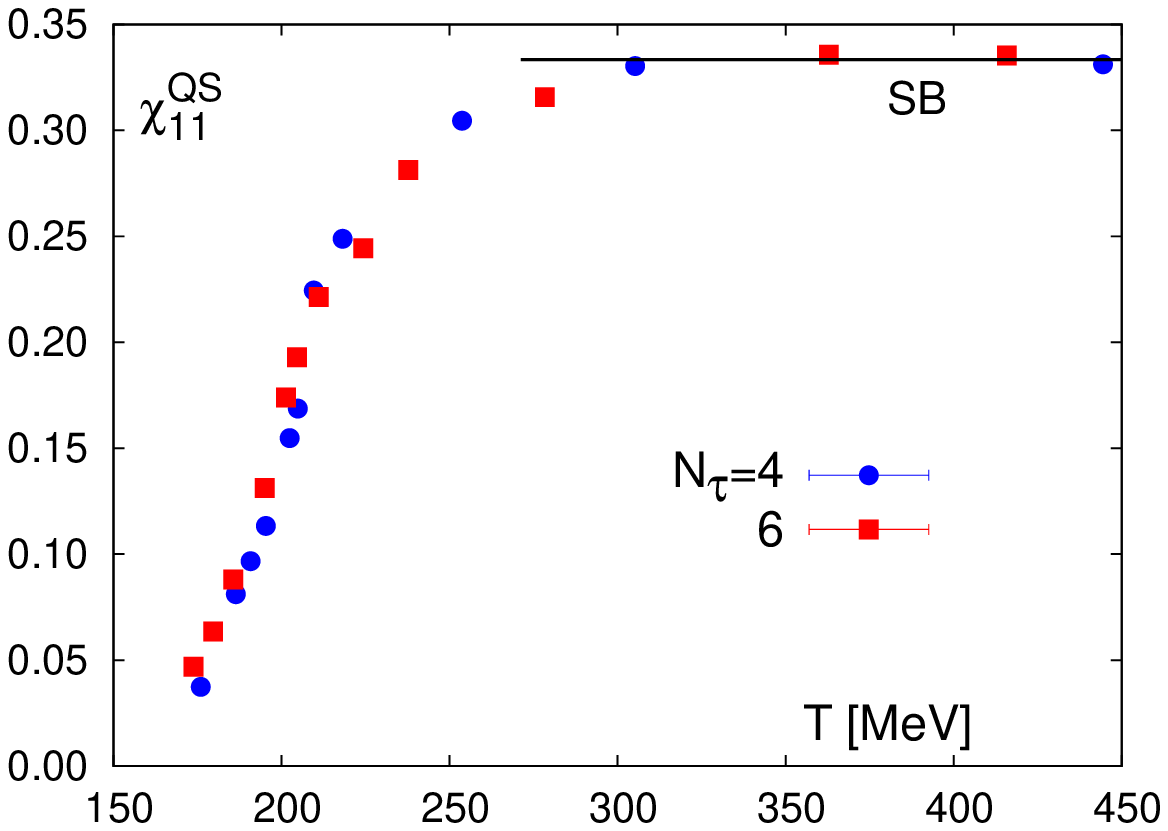, width=7.5cm}
\end{center}
\end{minipage}
\end{center}
\caption{Correlations of electric charge and strangeness (lower part) 
and the same quantity normalized to
quadratic fluctuations of electric charge (upper part). 
}
\label{fig:corQS}
\end{figure}

The analysis of cumulant ratios presented in the previous section
suggests that for temperatures above $1.5 T_c$ fluctuations of 
baryon number, strangeness and electric charge agree well with
the corresponding fluctuations in a non-interacting gas of  
light and strange quarks. In order to further test whether 
the relevant quasi-particle degrees of freedom indeed can be 
assigned to quarks, the analysis of correlations between different
charges is quite instructive \cite{vkoch,Gavai2}. Results for
correlations between $N_B$, $N_S$ and $N_Q$ are shown in 
Figs.~\ref{fig:cor} and \ref{fig:corQS}.
For completeness we also include the HRG-prediction. We note that the
model is not expected and, in fact, does not capture the fluctuations
and correlations in the transition region correctly. However, 
as shown in Figs. 6 and 7 as well as in Fig. 4 it reproduces
qualitatively the features of fluctuations and correlations. 
In particular, the drop and
rise seen in our numerical results for $BQ$ and $BS$ correlations,
respectively, is seen also in the HRG model calculations. To put these
observations on a more quantitative basis will require further
calculations at lower temperatures.

In Fig.~\ref{fig:cor} the correlation functions in the numerator project
only on the charged baryon sector of the spectrum. 
The ratio $\chi_{11}^{BQ}/\chi_2^B$
therefore approaches\footnote{We ignore here small mass
differences of charged and neutral hadrons.} $1/2$ in the low temperature 
limit as the numerator 
receives contributions from protons and anti-protons only, while the 
denominator also receives contributions from the neutrons. 
The ratio $\chi_{11}^{BS}/\chi_2^B$, however, 
will approach zero in the low temperature limit as
the lightest baryons carry no strangeness! 

In Fig.~\ref{fig:corQS} we show the correlation among $Q$ and $S$ normalized 
to the quadratic fluctuations of electric charge. A similar ratio, where
strangeness fluctuations have been used to normalize the $Q$-$S$ 
correlations, has been discussed in \cite{vkoch} and has also been calculated 
in 2-flavor QCD with a quenched strange quark sector \cite{Gavai2}. We prefer the above
normalization, as it emphasizes the non trivial correlations between $Q$ and
$S$ that persist in the high temperature phase of QCD. Unlike $\chi_2^S$
the charge fluctuations $\chi_2^Q$
approach the ideal gas value rapidly above $T_c$ (see Fig.~\ref{fig:fluc}).
The deviations from ideal gas behavior seen in Fig.~\ref{fig:corQS}
thus are mainly due to the deviations of $\chi_{11}^{QS}$ from ideal 
gas behavior. This is shown in the lower part of Fig.~\ref{fig:corQS}.
Apparently, the temperature dependence of $\chi_{11}^{QS}$ is very 
similar to that of $\chi_2^S$.

\section{Conclusions}

We have analyzed the fluctuations of baryon number, electric charge and
strangeness in finite temperature QCD at vanishing chemical potential.
A comparison of calculations performed with ${\cal O}(a^2)$ improved 
staggered fermions for two different values of the lattice cut-off shows 
that these effects are generally small; cut-off effects as well as 
the dependence of observables on the light quark masses become, however, 
more severe for higher order cumulants. 

We find fluctuations and correlations of conserved charges are well 
described by an ideal, massless quark gas already for temperatures of
about ($1.5-1.7$) times the transition temperature. Deviations from
ideal gas behavior seem to be mainly induced by the strange quark
sector for which the quadratic quark number fluctuations 
approach the ideal gas limit more slowly. This cannot only be
explained by the quark mass dependence of an ideal gas but
seems to reflect a significant quark mass dependence of the interaction 
at high temperature. 

At low temperature we find that fluctuations and correlations of 
conserved charges are 
reasonably
well described by a hadron resonance gas 
up to temperatures close to the transition temperature. Higher 
order cumulants, however, signal that the resonance gas prescription
has to break down at temperature values close but below $T_c$. 

The current analysis has been performed with light quarks that are 
one tenth of the strange quark mass. We have shown, that higher order
cumulant ratios like, for instance $\chi_6^Q/\chi_2^Q$, become
quite sensitive to the pseudo-scalar Goldstone mass. In numerical
calculations with staggered fermions this also means that results
become sensitive to a correct representation of the entire Goldstone
multiplet. Calculations with smaller quark masses closer to the 
continuum limit will thus be needed in the future to correctly 
resolve these higher order cumulants, which will give deeper insight
into the range of applicability of the resonance gas model at low
temperature and the non-perturbative features of the QGP above but
close to $T_c$.

\section*{Acknowledgments}
\label{ackn}
This work has been supported in part by contracts DE-AC02-98CH10886
and DE-FG02-92ER40699 with the U.S. Department of Energy,
the Bundesministerium f\'ur Bildung und Forschung under grant
06BI401, the Gesellschaft
f\"ur Schwerionenforschung under grant BILAER and the Deutsche
Forschungsgemeinschaft under grant GRK 881. Numerical simulations have
been performed
on the QCDOC computer of the RIKEN-BNL research center, the DOE funded
QCDOC at BNL, the apeNEXT at Bielefeld University and the BlueGene/L
at the New York Center for Computational Sciences (NYCCS). 

\vspace*{0.2cm}
\appendix*
\section{Quark mass dependence of cut-off effects}

Cut-off effects arising from the introduction of a finite lattice
spacing and, in particular, their dependence on the discretization
scheme have been extensively discussed in the ideal gas limit.
However, standard as well as ${\cal O}(a^2)$ improved staggered 
fermion formulations have usually been analyzed in the massless 
limit, which is appropriate for the light quark sector of QCD.
For the strange quark sector non-zero mass effects may play a 
role, in particular close to the QCD transition temperature which
is only about twice as large as the strange quark mass. We present
here an analysis of the quark mass dependence of cut-off effects in
the ideal gas limit.

The quark number susceptibilities are obtained as the second
derivative of the pressure, $p/T^4$, with respect to the
chemical potential normalized to the temperature, $\mu/T$.
An analysis of the cut-off and quark mass dependence of 
$\chi_2^{u,s}$ thus can follow closely the corresponding
analysis performed for the pressure\footnote{Note that at 
vanishing chemical potential the pressure is simply related to 
the free energy density, $p=-f$.} in the massless limit
\cite{cutoff}. On a lattice of size $N_\sigma^3N_\tau$ 
this is given by 
\begin{equation}
\frac{p}{T^4} = 2 \left(\frac{N_\tau}{N_\sigma}\right)^3 
\sum_{\mathbf{p},p_4} \ln (D^2 +m^2),
\end{equation}
with, $D^2 = \sum_{\mu=1}^4 D_\mu^2$ where
\begin{eqnarray}
D_1 = &&\hspace*{-0.3cm}c_{10}\sin(p_1)+c_{30}\sin(3p_1)+2c_{12}\sin(p_1)\times\nonumber \\
&&\hspace*{-0.3cm}\left[\cos(2p_2)+
\cos(2p_3)+\cos(2p_4)\right]
\end{eqnarray}
and $D_2$, $D_3$ and $D_4$ are obtained by cyclic permutation of the $p_i$s.
For the standard (naive) discretization scheme, one has 
$c_{10}=1$, $c_{30}=c_{12}=0$; 
the discretization scheme used in the p4-action corresponds to
$c_{10}=3/4$, $c_{30}=0$, $c_{12}=1/24$ and for the Naik action it is
 $c_{10}=9/8$ and $c_{30}=-1/24$ and $c_{12}=0$.

\begin{figure}[t]
\begin{center}
\begin{minipage}[c]{7.5cm}
\begin{center}
\epsfig{file=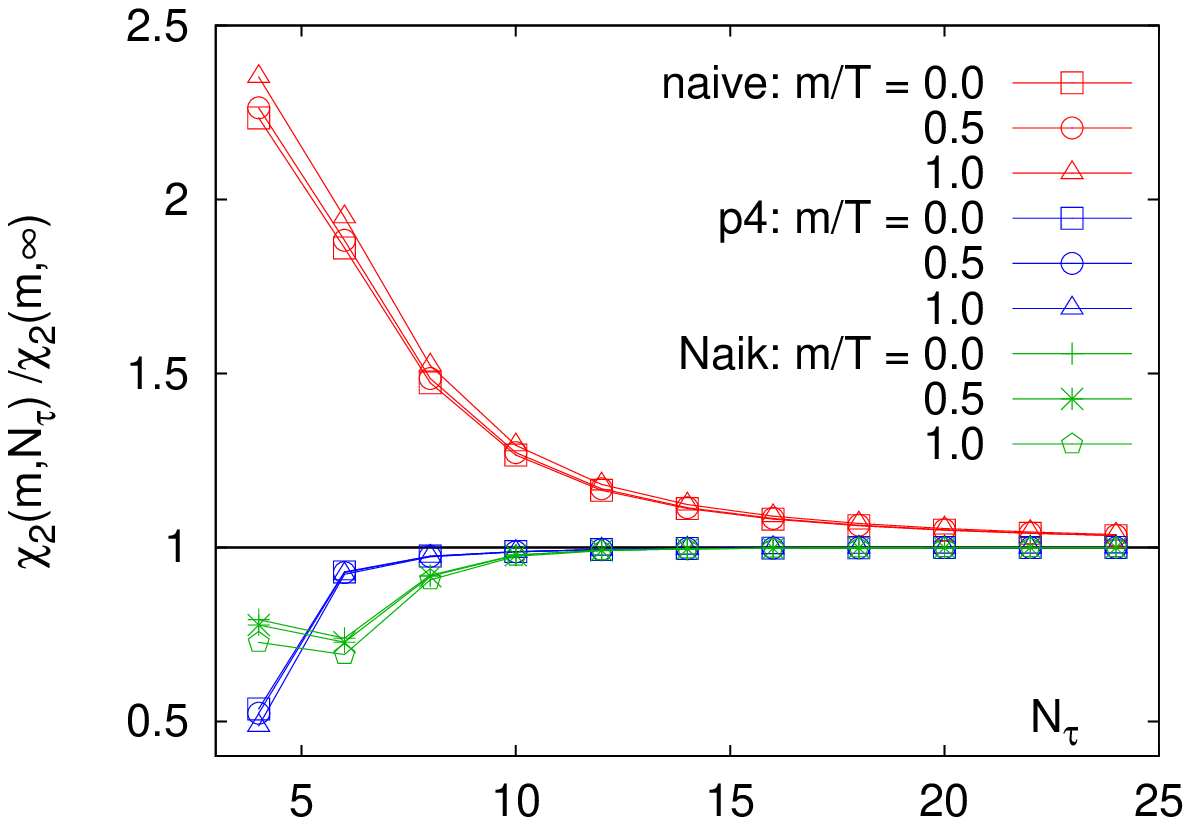, width=7.5cm}
\epsfig{file=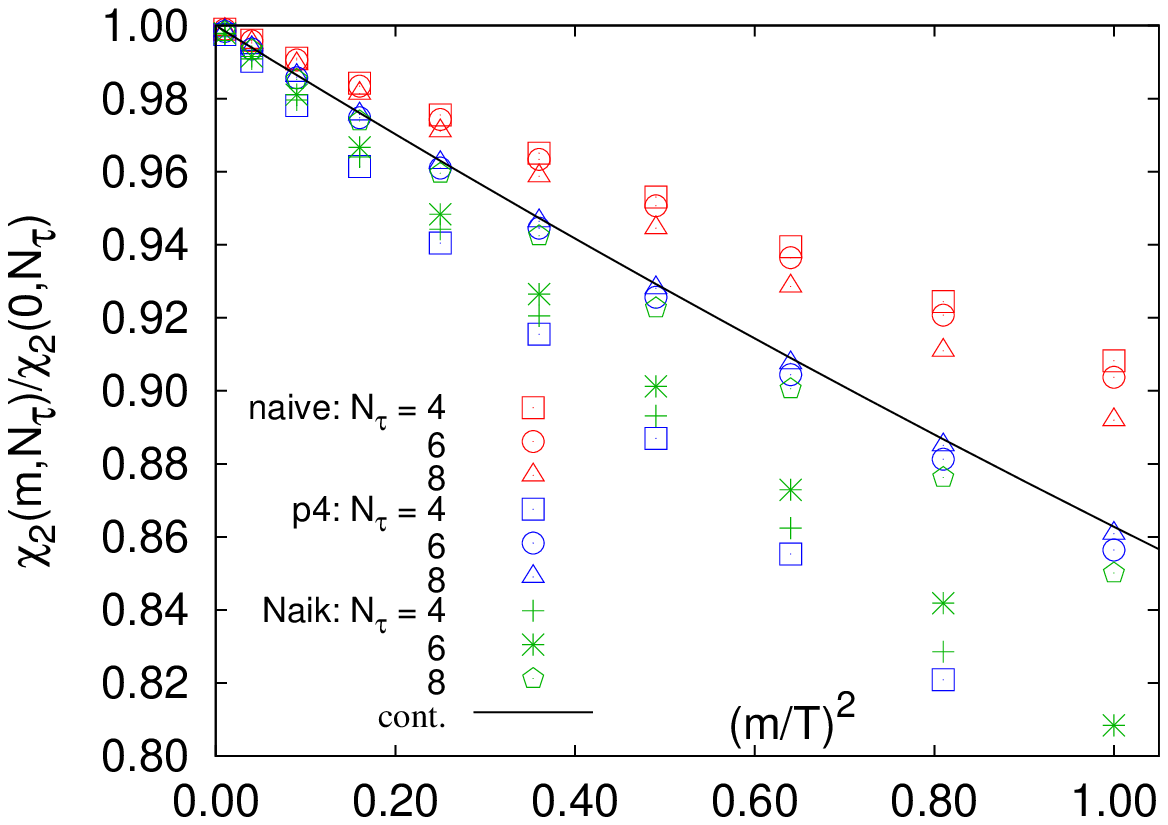, width=7.5cm}
\end{center}
\end{minipage}
\end{center}
\caption{Cut-off dependence of quark 
number fluctuations in the infinite
temperature, ideal gas limit. Shown
are results for staggered fermions in the
standard (naive), p4 and Naik discretization scheme.
}
\label{fig:SB}
\end{figure}

We evaluate here quark number susceptibilities as a function of quark 
mass on lattices with finite temporal extent
$N_\tau$ and infinite spatial volume for the
standard (naive) staggered fermion discretization and 
the improved (p4) discretization scheme. For vanishing
quark mass results on the cut-off dependence of 
quark number susceptibilities in the
ideal gas limit have also been presented in Ref.~\cite{gavaicut,cutoff1}.
There also some results for $\chi_4$ and $\chi_6$ can be found. 

Starting with
the expressions for $p/T^4$ at non-zero chemical potential
($p_4\rightarrow p_4-i \mu$)
and taking two derivatives with respect to $\mu/T$ one 
finds in the standard scheme for $\mu \equiv 0$,
\begin{align}
\begin{split}
\chi_2(m,N_\tau) &= 2\left(\frac{N_\tau}{N_\sigma}\right)^3
\sum_{\mathbf{p},p_4}(A - B + AB), \\
 A &= \frac{2 \sin^2(p_4)}{m^2 + \sum_{i=1}^4 \sin^2(p_i)},\\
 B &= \frac{2 \cos^2(p_4)}{m^2 + \sum_{i=1}^4 \sin^2(p_i)}. 
\label{eq:chi2_naive}
\end{split}
\end{align}
the corresponding expressions for the p4 and Naik actions are 
somewhat more complicated but can be derived straightforwardly.

In the continuum limit the quark number susceptibility
is given by,
\begin{equation}
\chi_2(m/T, \infty) = \frac{6}{T^3\pi^2}\int_0^{\infty} {\rm d}k
\frac{k^2\; {\rm e}^{-E/T}}{\left( 1+ {\rm e}^{-E/T}\right)^2}
\; , 
\label{continuum}
\end{equation}
with $E^2 = k^2+m^2$.
We note that for small values of the quark mass $\chi_2$
is quadratic in the mass, 
\begin{equation}
\chi_2(m/T, \infty) = 
1-  \frac{3}{2\pi^2}\left( \frac{m}{T}\right)^2 +{\cal O}((m/T)^4)\; .
\label{expand}
\end{equation}

We have evaluated $\chi_2(m,N_\tau)$ for various values
of $N_\tau$ and fixed quark mass as well as for fixed
$N_\tau$ varying the quark mass. The results for the
naive and p4 actions are shown in Fig.~\ref{fig:SB}

As expected, the cut-off dependence of $\chi_2$ at
non-zero values of the quark mass
closely follow the pattern known from the analysis
of the pressure at finite $N_\tau$ and $m/T=0$.
While the standard discretization scheme shows large
deviations from the continuum result even for 
$N_\tau \sim 10$, the results for the p4 action
come close to the continuum result already for 
$N_\tau =6$. This also is true for the ratio
$\chi_2(m,N_\tau)/\chi_2(0,N_\tau)$. Although in 
this ratio a large part of the cut-off dependence
cancels, the quark mass dependent corrections 
remain. 

We note, that for finite $N_\tau$ the quark mass
dependence of $\chi_2(m,N_\tau)/\chi_2(0,N_\tau)$
in the naive discretization scheme is weaker than
in the continuum limit. This is qualitatively
different from the differences seen in the lower
part of Fig.~\ref{fig:c2uc2s} between the standard 
and improved discretization schemes.

\end{document}